\documentclass{PoS}

\title{Coarse graining hadronic scattering}

\ShortTitle{Coarse graining hadronic scattering}

\author{\speaker{Enrique Ruiz Arriola}\thanks{This work is partly
      supported by the Spanish Ministerio de Econom\'{\i}a y
      Competitividad and European FEDER funds (grant
      FIS2017-85053-C2-1-P), Junta de Andaluc\'{\i}a grant FQM-225,
      the Swiss National Science Foundation and the {\bf Craven Allen
        Gallery--House of Frames, Durham (NC)}}\\ Departamento de
  F\'{\i}sica At\'omica, Molecular y Nuclear and Instituto Carlos I de
  Fisica Te\'orica y Computacional, Universidad de Granada, E-18071
  Granada, Spain\\ E-mail: \email{earriola@ugr.es}} \author{Jacobo
  Ruiz de Elvira \\ Albert Einstein Center for Fundamental Physics,
  Institute for Theoretical Physics, \\ University of Bern,
  Sidlerstrasse 5, CH--3012 Bern,
  Switzerland. \\  E-mail:\email{elvira@itp.unibe.ch}}

\abstract{We show that it makes sense to coarse grain hadronic
interactions such as $\pi\pi$ and $\pi N$ reactions following
  previous work on NN scattering. Moreover, if the interaction is
  taken to be given by chiral dynamics at long distances above a given
  value $r > r_c$ larger than the elementary radii of the interaction
  hadrons the unknown short distance region $r< r_c$ is characterized
  by a {\it finite} number of fitting parameters. This number of
  independent parameters needed for a presumably complete description
  of scattering data for a CM energy below $\sqrt{s}$ has been found
  to be given by $N_{\rm Par} = N_S \times N_I \times (p r_c )^2 /2 $
  with $N_S$ and $N_I$ the number of spin and isospin channels, and
  $p$ the CM momentum respectively.  Therefore, for an experiment (or
  sets of experiments) with a total number of data $N_{\rm Dat}$ the
  number of degrees of freedom involved in a $\chi^2$-fit is given by
  $\nu = N_{\rm Dat}-N_{\rm Par}$ and confidence levels can be
  obtained accordingly by standard means.  Namely a $1 \sigma$
  confidence level corresponds to $\chi_{\rm min}^2/\nu \in (1-
  \sqrt{2/\nu},1+\sqrt{2/\nu})$. We discuss the approach for $\pi\pi$
  and $\pi N$ with an eye put on a data selection program and the
  eventual validation of chiral symmetry.}

\FullConference{The 9th International workshop on Chiral Dynamics\\
		17-21 September 2018\\
		Durham, NC, USA}

\begin{document}

\section{Introduction}

The effective field theory (EFT) approach advocated by Weinberg 40
years ago~\cite{Weinberg:1978kz} has provided a framework where the
consequences of chiral dynamics can be best exploited quantitatively
via Chiral Perturbation Theory (ChPT)(see e.g. \cite{Weinberg:2016kyd}
for un upgraded view). The characteristic feature of the approach is
the declaration of a power counting within a prescribed range of
validity and the subsequent proliferation of parameters in
perturbation theory consistently with relativity, analiticity,
unitarity and crossing. The validation of an EFT against experimental
data with given uncertainties requires a comparable theoretical
precision due to the finite truncation of the perturbative
expansion. Clearly both too noisy theory or experiment cannot be
falsified. The usual situation in hadronic physics has always been
that the theory has larger uncertainties than experiment, so that we
should expect that precise experiments may need a huge number of EFT
parameters. In this contribution we want to review a less fundamental
but general approach which allows one to analyze scattering data with
the number of parameters being fixed {\it independently} of the
experimental precision.

Scattering experiments in hadronic physics involve a given maximal CM
energy $\sqrt{s}$. Hence, they provide a maximum achievable resolution
$\Delta r= \hbar / p_{\rm CM}$ corresponding to the minimal de Broglie
wavelength beyond which no information can be obtained. Thus, an
interaction of range $a$ is effectively sampled at a {\it finite}
number of points, $ N \sim a/\Delta r $ which, in the absence of
further information, one may assume that it provides a finite number
of {\it independent} strength parameters. From a phenomenological
perspective these parameters can be used as fitting variables to
describe experimental data {\it below} the maximal $\sqrt{s}$.  We
expect a good fit quality, {\it regardless} of the accuracy of the
experimental data. This is the essence of the coarse grained
interactions, which ultimately implement the Wilsonian renormalization
point of view in configuration space. The advantage over the momentum
space formulation is that i) long distance interactions are given by
particle exchange and hence are local at sufficiently large distances
and ii) the implementation of Coulomb effects necessary to describe
scattering data of charged particles is straightforward. The present
contribution reviews and ellaborates on previous studies about $NN$,
$\pi\pi$ and $\pi N$ interactions.

Thus, under the assumption that all (published) experiments are
correct, a viable and possibly optimal compromise is to validate the
largest possible number of data by using a theoretical model that
congregrates as many data as possible so that it checks for
statistical consistency as a whole. Following previous positive
experience, we review the coarse graining approach to arbitrate and
discriminate between the possible differences. Once we start
discarding data as possible outliers there is no way to disentangle
possible restrictions imposed by the theoretical model from genuinely
real inconsistencies. This said, the figure of merit corresponds to
maximize the number of data for a given confidence level. Clearly, any
analysis will be updated by including more {\it consistent} data than
the previous analysis with a comparable fit quality.

\section{Inconsistent Data}

The above expectations should hold regardless on the precision of
experimental data. However, this implicitly assumes not only correct
measurements with removed systematic uncertainties but also that the
statistical data uncertainties have correctly been estimated. Besides,
when experiments are conducted under a variety of conditions or in
different labs the issue of consistency becomes pertinent. One should
realize that any experiment may appear correct if error bars are
sufficiently large.  The issue may become critical since different
experiments may have their own source of bias and experimentalists may
assume a completely unrealistic uncertainty; at the extreme they may
believe that either they have benchmarking precision (because there is
a reward for precision) or that they make very conservative estimates
(because they do not want to be wrong). Both extremes are undesirable
and may lead ultimately to reject otherwise valuable and perhaps
costly data, since they either become unduly influencial (a too high
contribution the $\chi^2$ ) or effectively irrelevant ( a too small
contribution to the $\chi^2$). While the abundance of data is usually
perceived as an increase on the statistics, and hence an improvement
on the precision, it often has the side effect that many more
inconsistences may also arise and a data selection process becomes
mandatory. Fortunately, the large number of experiments suggests that
this selection may be implemented by a majority vote type of
argument~\cite{RuizArriola:2017kqs}.

\begin{figure}
\centering
\includegraphics[scale=.8]{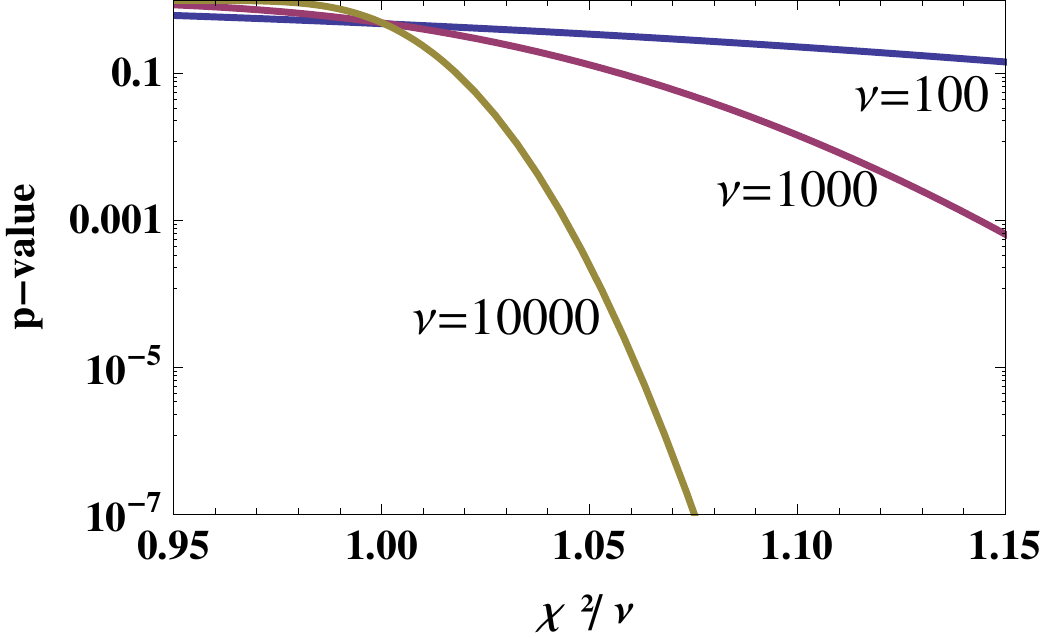}
\caption{Colour online: The p-value as a function
  of the reduced $\chi^2$ for different number of degrees of freedom
  $\nu=100,1000,10000$.}\label{Fig:pValue}
\end{figure}

The simplest hadronic reaction is elastic $\pi\pi$ scattering. From
the theoretical side it is constrained by relativity, unitarity,
causality, crossing and eventually chiral symmetry provide a practical
framework. Unfortunately, it cannot be directly observed
experimentally, only through pion production experiments. Actually,
most of the analysis is dominated by systematic errors, so the
question on data selection is subtle (see \cite{Perez:2015pea}). In
contrast, $NN$ and $\pi N$, are directly measurable. The largest
database to date is provided by the SAID analysis which is a PWA up to
a maximum energy $\sim 3 \, {\rm GeV}$. The corresponding number of
data and minimal $\chi^2$ values are reviewed in Table \ref{tab:said}.
The poor quality of the fits is evident from the tiny p-values. We
remind that it corresponds to the probability for a given statistical
model when the null hypothesis is true. In Fig.~\ref{Fig:pValue} the
situation is vividly illustrated, where the often used rule of
$\chi^2/\nu \sim 1$ for a good fit is wrong for $\nu > 100$ The
situation of Table \ref{tab:said} may be interpreted as a signal for a
faulty model or mutually inconsistent data or both.

As it is well known, when two experimental data are directly compared
for the same input variables and found to be inconsistent within
uncertainties one faces the problem that either one of them or both
are necessarily wrong. A more subtle issue is when the input variables
are not exactly the same, since a direct comparison is not possible
and a discrepancy might also well be a significant signal. If there is
a priori no reason for this signal, and one assumes smoothness in the
input variables, some interpolation or extrapolation becomes possible
and one can then address the issue of consistency. Of course, the best
possible situation occurss when this smoothnes assumption is
prescribed by theory. Thus, what theory might one use which can still
be right but flexible enough to accomodate as many data as possible ?.

\begin{table}
\begin{center}
  \begin{tabular}{ l|ccccr }
   & $T_{\rm LAB}$ (MeV) & $N_{\rm Dat}$ & $\chi^2 $ & $\chi^2 /N_{\rm Dat}$ & p-value \\
    \hline 
  pp & 3000 & 25188 & 48225.0 & 1.9 & $4 \times 10^{-1453}$\\
  np & 3000 & 12962 & 26079.9 & 2.0 & $1 \times 10^{-883}$\\
  $\pi N$ & 3000 & 41926 & 166585.05 & 4.1 & $4 \times 10^{-14513}$\\
   $\pi N$ & 300 & 2599 & 4586.2 & 1.8 &$4 \times 10^{-113}$ \\
\end{tabular}
\end{center}
\caption{$NN$ and $\pi N$ PWA characteristics from SAID database
  \sf{http://gwdac.phys.gwu.edu/}. The p-value is $ p= \Gamma
  \left(\nu/2,\chi^2/2 \right)/ \Gamma \left(\nu /2 \right)$ with
  $\Gamma(a,x)$ and $\Gamma(x)$ incomplete and complete Euler Gamma
  functions respectively. We also take $\nu \sim N_{\rm Dat}$ since
  $N_{\rm Par} \lll N_{\rm Dat}$}
\label{tab:said}
\end{table}

In hadronic physics, at least at low energies, there is a bunch of
field theoretical conditions involving relativity, analiticity,
unitarity, crossing and chiral symmetry that apply only to the strong
interactions, which are short range.  Are these conditions powerful
enough to discard some reported scattering data in favour of other
scattering data ?. A practical problem is related to the presence of
long range effects such as Coulomb interaction, vacuum polarization or
magnetic interactions, etc. which are hard to implement, say, within
dispersion relations, but influence peripheral scattering. The
standard approach is to first ``clean'' the database by other method
and then ``remove'' the long range effects from the data, so that the
theory can then be tested. A particularly revealing example is the
recent $np$ and $pp$ analysis of scattering data, where the so called
coarse grained potential model has been found to provide the largest
mutually $3 \sigma$ self-consistent Granada-2013 database at about
pion production threshold~\cite{Perez:2013jpa}.  One important
consequence of this database is that paves the road for theoretical
tests such as ChPT~\cite{RuizArriola:2016sbf}. In fact, it has
recently been shown~\cite{Reinert:2017usi} that ChPT for NN at N5LO
allows quite accurately to describe this database, in fact,
outperforming the previously popular potentials which had been
considered ``high quality''~\footnote{It is fair to say that despite
  these good features, strictly speaking the resulting p-value is
  still unsatisfactory and the declared power counting has to be
  suitably but inconsistently reshuffled. See
  also~\cite{Entem:2017gor}.  If ChPT had been used to {\it select} NN
  scattering data, many more data would have been rejected than in the
  Granada-2013 database.}.

\section{Invariant momentum approach}

From a fundamental point of view the relativistic two body problem is
best analyzed in terms of the honorable Bethe-Salpeter equation
(BSE)~\cite{Salpeter:1951sz}, where exact solutions and numerical
methods have been studied exhaustively over the years (for a review on
the relativistic few body problem see e.g. \cite{Carbonell:1998rj},
and references therein). This is a good approach as long as the input
of the equation, the two particle irreducible (2PI) 4-point funtion is
known {\it exactly}. However, one must point out that the 2PI Green
function is mostly accessible {\it only} within perturbation
theory. The problem with this approach is that for the usual hadronic
couplings the ultraviolet structure of the 2PI 4-point function
becomes increasingly divergent and thus the scattering problem becomes
ill defined. For instance, numerous studies have been concerned over
the years about minimal sets of diagrams which implement compelling
properties. However, in order to compare to experiment, the 2PI kernel
is ultimately phenomenologically parameterized and interpreted from a
fit to experimental data. Therefore many, in principle different, 3D
reductions are in fact equivalent from a perturbative point of
view~\cite{Nieves:1999bx}.

Besides, there is the problem of off-shellness; a perturbative
evaluation of the amplitude yields unique results only when all
variables are taken to be on-shell. Off-shell extrapolations are in a
sense arbitrary and also depend on the field
parameterization~\cite{Arriola:2016pnw}.  A fairly old but unknown
study~\cite{namyslowski1967relativistic} analyzes the interplay
between on-shell and off-shell in terms of a non-linear dispersion
relation (see e.g. Ref.~\cite{Lutz:2015lca}).  Under those conditions
it is far more useful to use a much simpler approach such as the
mass-invariant method~\cite{Allen:2000xy} which is based on rewritting
the problem and matching the potential perturbatively as has been
proposed e.g.  in Refs.~\cite{Arriola:2016pnw}. For the case of
unequal masses the mass-invariant scheme generates a complicated
equation, but we can instead proceed in a CM-momentum invariant
approach, namely
\begin{eqnarray}
  \left[ -\nabla^2 + U(\vec r) \right] \Psi= p_{\rm CM}^2 \Psi \, , 
  \label{eq:invp}
\end{eqnarray}
where for $W= \sqrt{s}= \sqrt{p^2+M^2}+\sqrt{p^2+m^2}$ (we drop the
subscript CM from now)
\begin{eqnarray}
p^2 = \frac{(m-M-W) (m+M-W) (m-M+W) (m+M+W)}{4 W^2} \, . 
\end{eqnarray}
The upshot is that for all purposes we can use a Schr\"odinger type
equation where the non-relativistic CM momentum is promoted to a
relativistic one as given by the previous equation where the
interaction $U(\vec r)$ can be determined by matching to ChPT at long
distances, $r > r_c$ and the coarse graining principle can be applied
to small distances $r < r_c$ with $r_c$ a suitable short distance
cut-off.

\section{Elastic scattering and finite resolution}

For two particle scattering the spatial resolution in the relative
cooredinate is fixed by the shortest relative de Broglie wavelength,
which is given by
\begin{eqnarray}
\Delta r=\frac{\hbar}{p}= \lambda_{\rm min}/(2\pi) \, . 
\end{eqnarray}
Quantum field theory requires particle exchange to account for
interactions which for hadrons have a maximal range about the Compton
wavelength of the lightest hadron and are ${\cal O}(e^{-r/\lambda_C})$
with $r$ the relative distance between hadrons.  That means $\lambda_C
= \hbar /m_\pi c = 1.4 \, {\rm fm}$ for One Pion Exchange (OPE) as it
is the case for $NN$ interactions and $ \lambda_C = \hbar / 2m_\pi c
=0.7 \, {\rm fm}$ for Two Pion Exchange (TPE) which corresponds to
$\pi N$ and $\pi\pi$ interactions. Because of the exponential fall-off
we take the range numerically $a \sim 2 \lambda_C$.

\begin{figure}
\centering
\includegraphics[scale=.7]{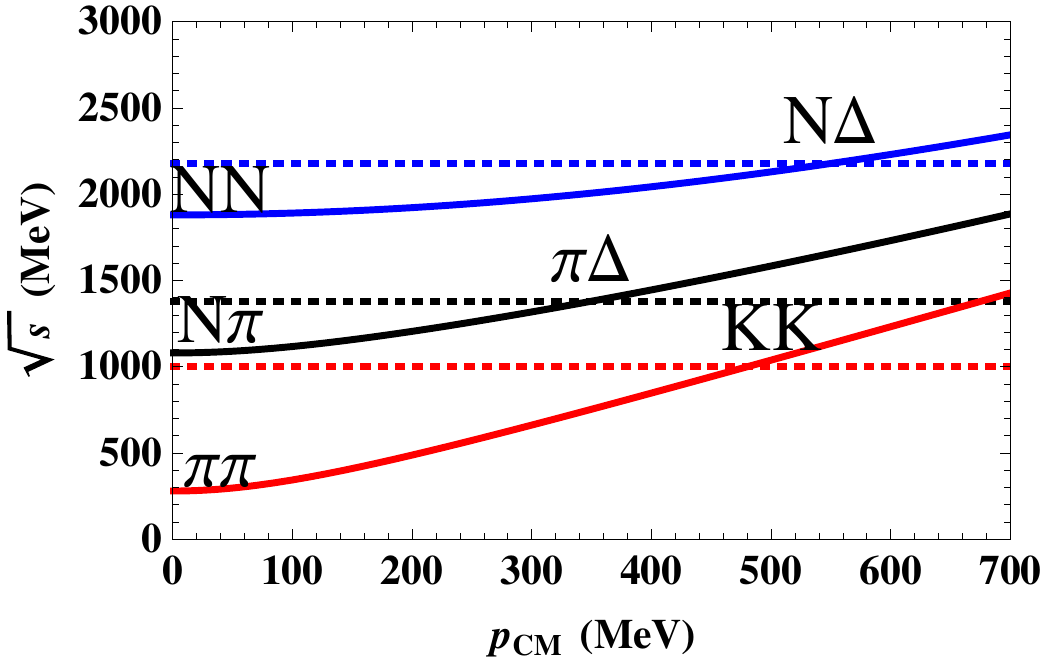}
\includegraphics[scale=.7]{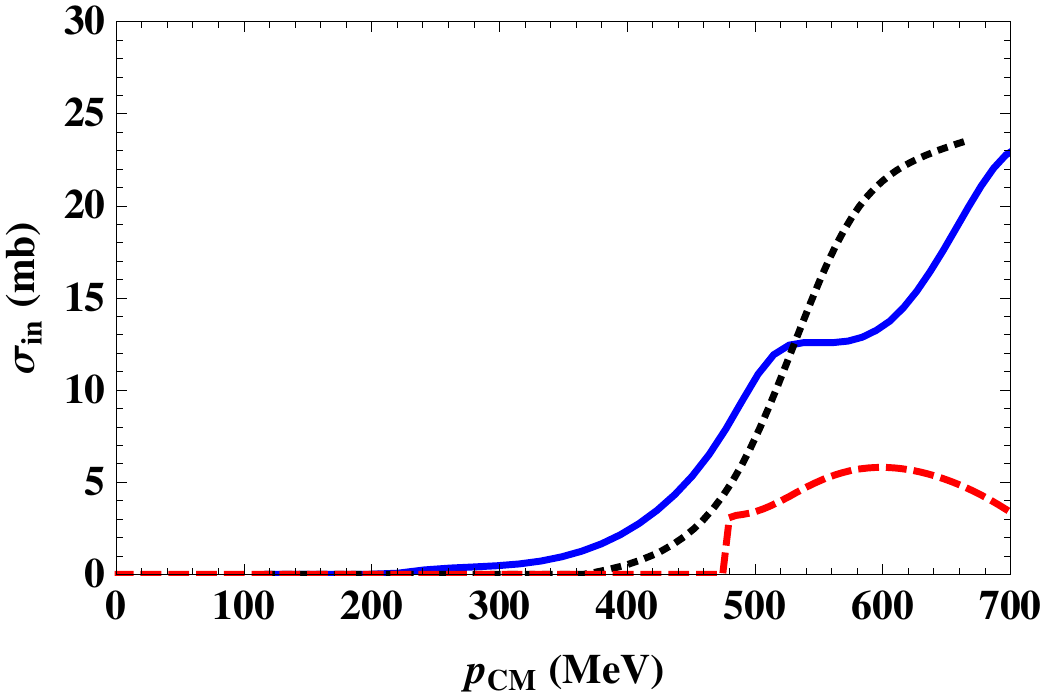} 
\caption{Colour online: (left panel) The CM energy as a function of
  the CM momentum for $\pi\pi$ (solid,red) , $\pi N$ (solid,black) and
  $NN$ (solid,blue).  We also plot the corresponding thresholds at
  $K\bar K$, $\pi \Delta$ and $N \Delta$ where the inelastic cross
  (right panel) section increases substantially (we take $\pi^0\pi^0$,
  $\pi^+ p$ and $pp$ respectively.}\label{Fig:VWpCM}
\end{figure}

In our presentation we will restrict to the case of elastic $\pi\pi$,
$\pi N$ and $NN$ interactions which definitely set an upper limit for
the CM energy. Inelastic thresholds are marked by pion production
processes, such as $\pi \pi \to \pi \pi \pi \pi$, $\pi N \to \pi \pi
N$ or $NN \to NN \pi$, but they generate small contributions to the
inelastic cross section so that their influence on the dominant
elastic process can be neglected. Actually, the inelastic cross
section experiences a rapid change when $\pi \pi \to K \bar K$, $\pi N
\to \pi \Delta $ and $ NN \to N \Delta$ corresponding to $\sqrt{s} = 2
m_K, m_\pi + M_\Delta , M_N + M_\Delta$ respectively.

In Fig.~\ref{Fig:VWpCM} we depict the CM energy as a function of the
CM momentum as well as the location of the thresholds where the
inelastic cross section undergoes a sudden rise. As we see this
happens in the CM momentum range of about a similar value $p_{CM} \sim
350-550$ which corresponds to a resolution $\Delta r = \hbar /p_{\rm
  CM} \sim 0.3-0.6 \, {\rm fm}$. As we see, they lie well at higher
energies than the single and double production thresholds. The reason
why the inelasticity is so tiny is likely to be found in chiral
symmetry and the fact that the pion couples derivatively, which
suppresses the production amplitude by extra powers of momentum. In
any case, these higher thresholds are the natural limit for a purely
elastic scattering description. Going to higher energies is possible
but it requires implementing energy dependence into the potential.

\section{Anatomy of hadronic interactions}

\subsection{Effective elementarity}

Hadrons have a finite size, which only becomes visible when the
probing wavelength is comparable. Longer wavelengths make the hadron
look as elementary and pointlike particles. This feature can be best
appreciated by analyzing the electromagnetic (em) interaction of charged
particles such as, e.g., $\pi^+ \pi^+ $, $\pi^+ p $ and $pp$. This can
be readily estimated by using
\begin{eqnarray}
  V_{AB}^{\rm em}(r) = \int d^3 x_1 \int d^3 x_2  \frac{\rho_A (\vec x_1)  \rho_B (\vec x_2)}{|\vec x_1  - \vec x_2 - \vec r|} \, , 
\end{eqnarray}
where $\rho_A(\vec x)$ and $\rho_B(\vec x)$ are the charge densities
corresponding to hadrons A and B respectively. Using the electric form
factors in the Breit frame
\begin{eqnarray}
 e F_{i}(\vec q) &=& \int    d^3 x \rho_{A,B} (\vec x)  e^{i \vec q \cdot \vec x} \, , 
\end{eqnarray}
with $i=A,B$, we obtain
\begin{eqnarray}
  V_{AB}^{\rm em}(r) = e^2 \int \frac{d^3 q}{(2\pi)^3} \frac{4 \pi}{\vec q^2} F_A
  (\vec q) F_B (\vec q) e^{i \vec q \cdot \vec r} \sim \frac{q_A q_B}{r} \qquad r \ge r_e \, . 
\end{eqnarray}
\begin{figure}
\centering
\includegraphics[scale=.8]{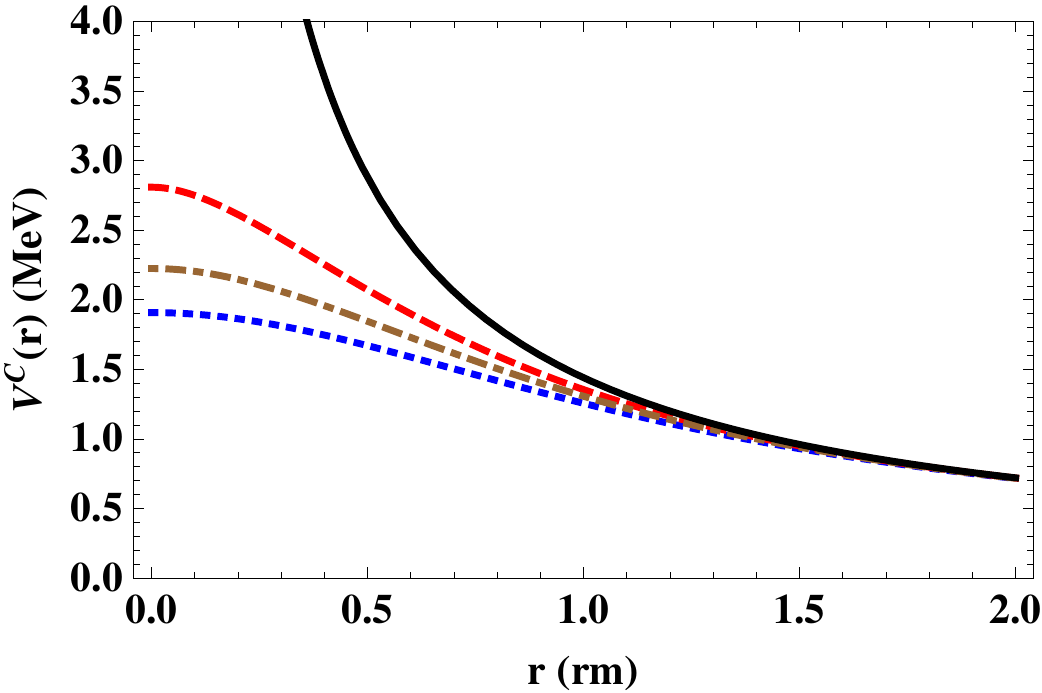} 
\caption{Colour online: Electromagnetic potentials for pointlike charge one particles (Solid,Black) and finite size particles , $\pi^+ \pi^+ $ (Red,dashed) ,
  $\pi^+ p $ (Brown, dotted-dashed) and $pp$ (Blue,dotted) as a function of the
  distance.}\label{Fig:Vem-all}
\end{figure}
As we see by looking at Fig.~\ref{Fig:Vem-all}, the elementarity
radius $r_e$ falls in the range $\sim 1.5 \, {\rm fm}$.  This suggests
that in order be entitled to ignore the finite hadron size, one should
take the cut-off distance to satisfy $r_c \ge r_e \sim 1.5 \, {\rm
  fm}$~\footnote{Here we take for simplicity the Vector Dominance form
  factor for the pion $F_\pi (\vec q) = m_\rho^2/(m_\rho^2+q^2)$ and
  the dipole form factor for the proton $F_p (\vec q)= 1/(1+\vec
  q^2/\Lambda^2)^2$ with $m_\rho=770 \, {\rm MeV}$ and $\Lambda=700 \, {\rm
    MeV}$.}. Of course, this effective elementarity feature occurs also for the
strong interaction, and we will assume that the corresponding
elementarity radius coincides with the one found in the em case.


\subsection{Impact parameter}

Due to angular momentum conservation one can use a partial wave
expansion for the quantum mechanical scattering amplitude, which in
the spinless case reads
\begin{eqnarray}
f(\theta,W)= \sum_{J=0}^\infty (2J+1) P_J(\cos \theta) \frac{e^{i \delta_J} \sin \delta_J}{p} \, . 
\end{eqnarray}
The standard semiclassical argument yields the relation 
\begin{eqnarray}
J+ \frac12 = p b \, ,  
\end{eqnarray}
between the impact parameter $b$ and the angular momentum $J$
The no-scattering situation corresponds to $b \ge a$ with $a$ the
range of the interaction, so that the partial wave expansion is
effectively truncated for a maximum angular momentum $J_{\rm max} +
1/2 \sim p a $. Operationally, one may take a more refined condition
where the phase-shift becomes zero within experimental uncertainties.
Actually, this is a model independent way of estimating {\it from} a
data analysis the finite range of the interaction $a$.

\begin{figure}
\centering
\includegraphics[scale=.8]{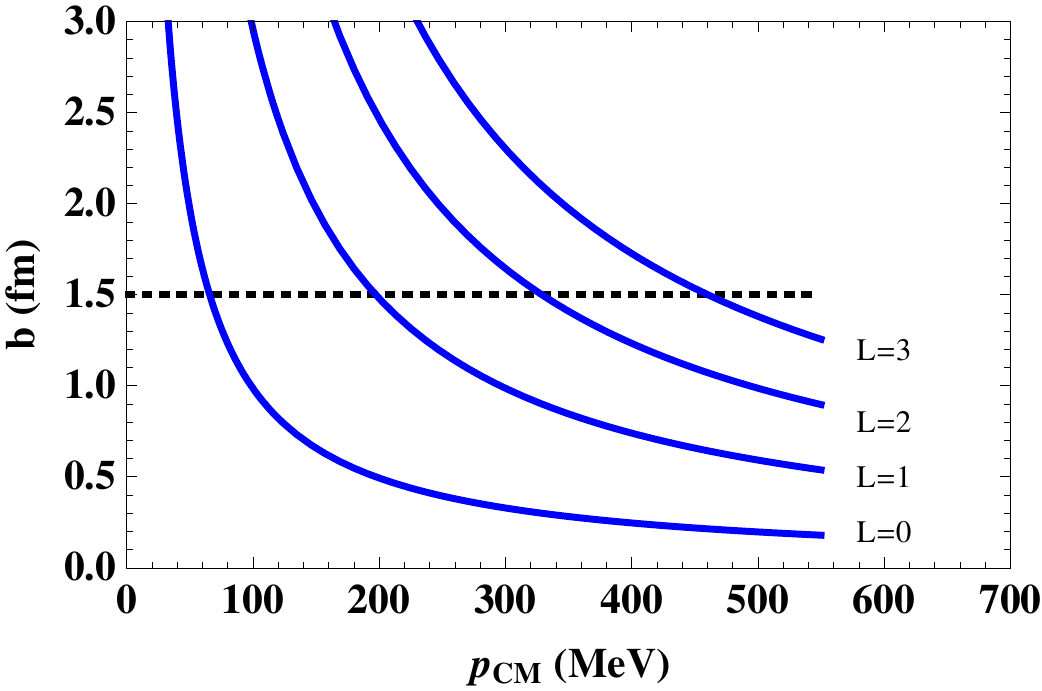} 
\caption{Colour online: The impact parameter as a function
  of the CM momentum for different partial waves with angular momenta
  $L=0,1,2,3$ (solid,blue). The (black,dotted) line indicates the
  critical radius marking the range of the interaction. 
}\label{Fig:Impact-pCM}
\end{figure}

\begin{figure}
\centering
\includegraphics[scale=.8]{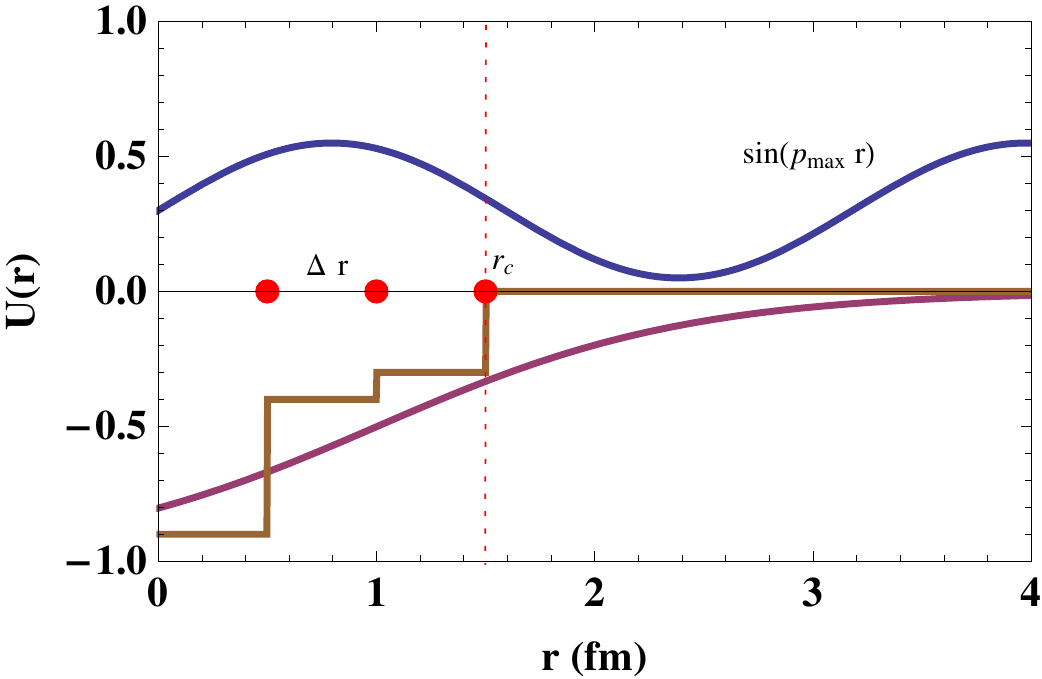} 
\caption{Colour online: An illustration of the coarse graining idea
  for a maximal CM momentum $p_{\rm max}=400 \, {\rm MeV}$ which
  corresponds to a wavelength $\lambda_{\rm min}=3$ and hence $\Delta r=0.5 \, {\rm fm}$
}\label{Fig:coarse}
\end{figure}

In Fig.~\ref{Fig:Impact-pCM} we show the dependence of the impact
parameter on the CM momentum for the lowest partial waves with angular
momentum $L=0,1,2,3$ assuming for simplicity $r_c=1.5 \, {\rm fm}$.
For instance, if we have $p_{\rm CM} \le 300 \, {\rm MeV}$ {\it only }
S- and P-waves probe the interaction region~\footnote{We are assuming
  a sharp boundary for the potential at $r=r_c$. If the potential
  $U(r)$is known {\it above} $r_c$, the argument applies equally,
  since roughly speaking peripheral waves with $J > J_{\rm max}$ are
  essentially determined from $U(r)$. A similar plot and further
  discussion about this may be found in
  Ref.~\cite{RuizSimo:2017anp}.}. For a given partial wave the idea of
coarse graining is illustrated in Fig.~\ref{Fig:coarse} for a maximal
CM momentum of $p_{\rm max}=400 \, {\rm MeV}$.

\begin{figure}
\centering
\includegraphics[width=.45\linewidth]{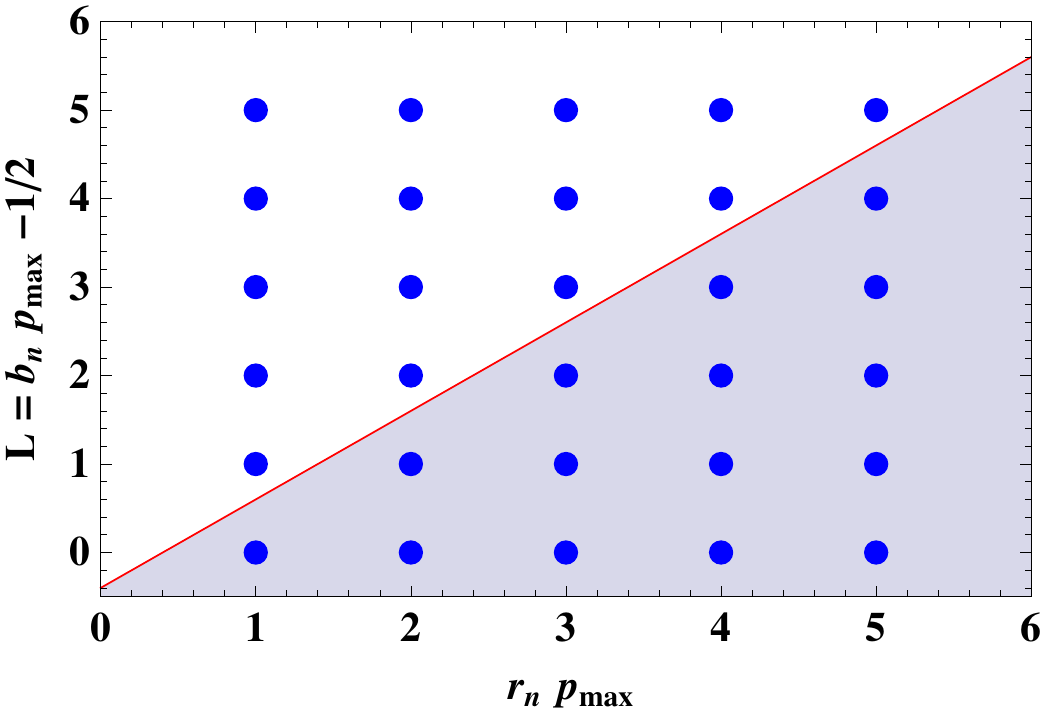}
\includegraphics[width=.45\linewidth]{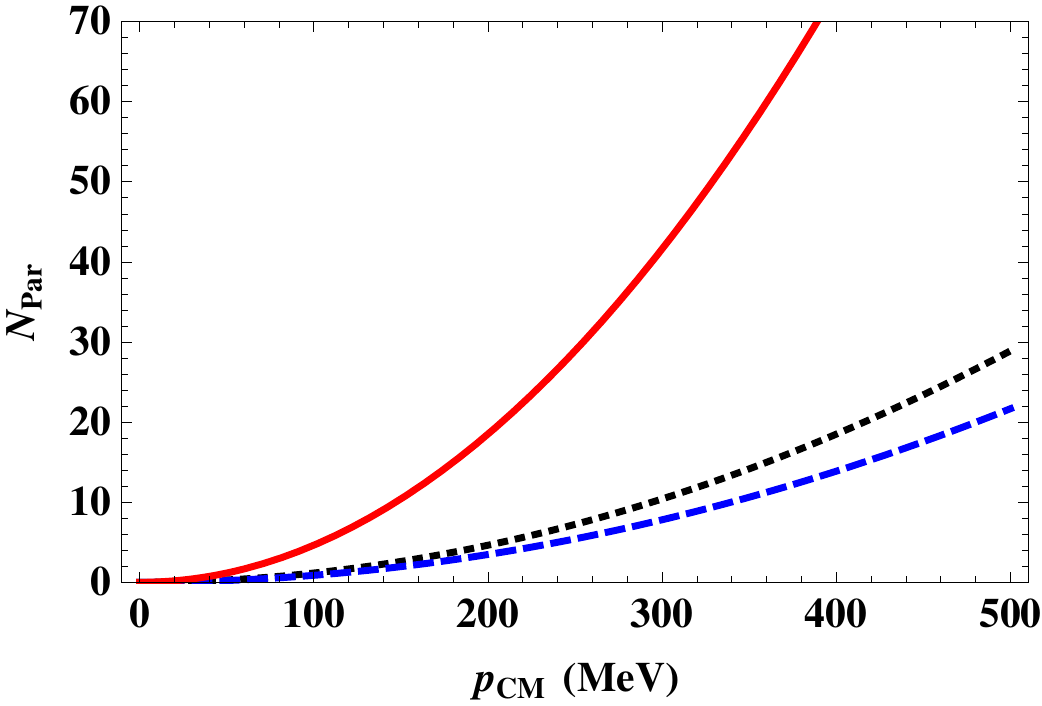} 
\caption{Colour online: (Left panel) Allowed independent parameters
  (shaded area) based on coarse graining the interaction below
  $r_c=1.5 \, {\rm rm}$. (Right panel) The estimated number of
  independent parameters as a function of the CM momentum for $\pi\pi$
  (dashed,blue) , $\pi N$ (dotted,black) and $NN$ (solid,red)
  scattering.}\label{Fig:Npar}
\end{figure} 

\subsection{The number of parameters}

Once we fix maximum CM momentum $p_{\rm CM, max}$ and the cut-off radius
$r_c$ the number of parameters $N_{\rm Par}$ can be easily calculated as follows.
For a central potential $U(r)$ one has a tower of angular momentum states
$L=0,1,2, \cdots$. The maximal angular momentum corresponds to 
\begin{eqnarray}
L_{\rm max} + \frac12 = p r_c
\end{eqnarray}
If we take the $L=0$ state, we have the grid points 
\begin{eqnarray}
\Delta r = \hbar /p_{\rm CM,max}  \to r_n = n \Delta r \, .
\end{eqnarray}
The idea is to use $U_n \equiv U(r_n)$ as fitting parameter.
According to the previous discussion, the number of points would be $
(r_c / \Delta r) \times L_{\rm max} \sim (p_{\rm CM,max} r_c )^2
$. However, the actual number is smaller since for $L \ge 1 $ some of
the points located at $r_n$ fall well below the centrifugal barrier,
and correspond to the classicaly forbidden region, thus their
contribution is suppressed. Taking into account both spin and isospin
degrees of freedom the following estimate can be obtained
as~\cite{Fernandez-Soler:2017kfu}
\begin{eqnarray}
N_{\rm par} = N_S N_I \sum_{L=0}^{L_{\rm max}} \sum_n \theta \left[ p^2-
  \frac{L(L+1)}{r_n^2}   \right] \sim \frac12 (p r_c)^2 N_S N_I  \, , 
\end{eqnarray}
where $N_S$ and $N_I$ the number of spin and isospin channels
respectively.  The number of independent parameters is plotted in
Fig.~\ref{Fig:Npar} as a function of the maximal CM momentum and for
$r_c=1.5 \, {\rm fm}$. Remarkably, for the $NN$ case, $N_{\rm Par}
\sim 50$ which explains the traditional number of fitting parameters
for ``high quality'' NN analyses over the years~\cite{Perez:2013cza}.
For $\pi \pi$ scattering below $K \bar K$ threshold, one needs $N_{\rm
  Par} \sim 9= 3 \times (S_{00}) + 3 \times (S_{20})+ 3 \times
(P_{11})$~\cite{RuizdeElvira:2018hsv}. For $\pi N $ scattering with
$\sqrt{s} \le m_\pi + M_\Delta = 1.38 \, {\rm GeV}$ the total number
of fitting parameters would be about $18= 3 \times (S_{11}+S_{31}) + 3
\times (P_{11}+P_{31}+P_{13}+P_{33}) $~\cite{Alarcon2019}.

\subsection{Complete potential}

The current attempts to verify the suitability of the coarse graining
approach have proved successful so far. The delta-shells approach is
by far the simplest scheme where the coarse graining of the
interaction can be implemented. The complete fitting potential (delta-shells)
\begin{eqnarray}
U (r) = \underbrace{\left\{ \sum_i \Delta r U(r_i) \delta (r -r_i) \right\}}_{\rm Short (Coarse \, grained)} \theta (r_e -r )+
\underbrace{U_{\rm ChPT}(r)}_{\rm Long (Particle \, exchange)} \theta (r-r_e) \, . 
\end{eqnarray}
Here $U_{\rm ChPT}(r)$ is the chiral part which can be obtained from
the corresponding Feynman diagrams having a $t$-channel
discontinuity~\cite{RuizdeElvira:2018hsv}. For $r_c \sim 1.5 \, {\rm
  fm}$ one takes $U_{\rm QFT}(r) = U_{\rm 1\pi}(r)+ U_{\rm 2\pi}(r)$ (
$1\pi$ and $2\pi$ exchange) for NN~\cite{Perez:2013oba,Perez:2014bua}
and $U_{\rm QFT}(r) =U_{\rm 2\pi}(r) $ ($2\pi$ exchange) for $\pi\pi
$~\cite{RuizdeElvira:2018hsv} and $\pi N$~\cite{Alarcon2019}.

\section{Closing remarks}

As we have discussed, for a maximal CM energy the number of
independent parameters characterizing the phenomenological interaction
is finite and in principle independent on the precision of the
experimental data.  A this point it is pertinent to compare with ChPT.
A simple way to estimate how many parameters are needed, proceeds by
analysing the scattering amplitude close to threshold. In the case of
$\pi\pi$ scattering and assuming isospin invariance, the scattering
amplitudes for $I=0,1,2$ depend on a unique function $A(s,t,u)$ which
for on-shell particles becomes $F(s,t)$. In the chiral counting and
going to small energies one has $s,t,m_\pi^2 = {\cal O} (p^2)$ we
expand
\begin{eqnarray}
F(s,t,m_\pi^2) = \sum_{k,l=0} \frac{c_{kl}^N}{f^N} s^k t^l (m_\pi^2)^{N-k-l} \, , 
\end{eqnarray}
with $f$ the pion weak decay constant. The number of independent
parameters, $c_{kl}^N$, corresponds to the number of partitions of the
order $N$ by three positive integers which is
$(N+1)(N+2)/2=3,6,10,15,21 \dots$ for $N=1,2,3,4,5, \dots $.  Thus,
ChPT has a finite applicability range with an increasing number of
parameters which improve the accuracy at a given finite order.
Similar features apply to $NN$~\cite{Perez:2013cza} and $\pi
N$~\cite{Alarcon2019}. Thus, the validation of the theory is related
to the accuracy of the {\it data}. Standard $\chi^2$ minimization
allows to determine the LEC's
\begin{eqnarray}
\chi_{\rm min}^2= \min_{c_1, c_2, \dots } \chi^2 (c_1, c_2, \dots)= \sum_{i=1}^{N_{\rm Dat}} \left[ \frac{{\cal O}_i^{\th} (c_1, c_2, \dots ) -{\cal O}_i^{\rm exp}}{\Delta {\cal O}_i^{\rm exp}} \right]^2
\end{eqnarray}
The problem is that we need $N_{\rm Par} \to \infty$ when $\Delta
{\cal O}^{\rm exp} \to 0 $ or $N_{\rm Dat} \to \infty$~\footnote{Of
  course, this is only valid around the threshold region.  Going to
  the resonance region is another story, as although many proposals
  implementing unitarity have been made, there is no uniquely and
  accepted method.}. In contrast, in the coarse graining approach we
expect $N_{\rm Par} \sim (p r_c)^2 $ regardless on $\Delta {\cal
  O}^{\rm exp} $. Whether or not the coarse graining approach
qualifies as expected for fitting and selecting, for instance, $\pi N$
scattering data remains to be seen.

 \section*{Acknowledgments}
{\it We  thank Rodrigo  Navarro P\'erez, Jose  Enrique Amaro  and Jose
  Manuel Alarc\'on for  discussions. One of us  (E.R.A.)  warmly thanks
  Keith Allen  Wenger and John  Craven Bloedorn for local  hosting and
  hospitality sponsored  by the Craven Allen  Gallery--House Of Frames
  (Durham, NC) and the stimulating atmosphere.}


\begin{thebibliography}{10}

\bibitem{Weinberg:1978kz}
S.~Weinberg, \emph{{Phenomenological Lagrangians}},
  \href{https://doi.org/10.1016/0378-4371(79)90223-1}{\emph{Physica} {\bfseries
  A96} (1979) 327--340}.

\bibitem{Weinberg:2016kyd}
S.~Weinberg, \emph{{Effective field theory, past and future}},
  \href{https://doi.org/10.1142/S0217751X16300076}{\emph{Int. J. Mod. Phys.}
  {\bfseries A31} (2016) 1630007}.

\bibitem{RuizArriola:2017kqs}
E.~Ruiz~Arriola, J.~E. Amaro and R.~Navarro~P\'erez, \emph{{Fitting and
  selecting scattering data}},
  \href{https://doi.org/10.22323/1.310.0134}{\emph{PoS} {\bfseries Hadron2017}
  (2018) 134}, [\href{https://arxiv.org/abs/1711.11338}{{\ttfamily
  1711.11338}}].

\bibitem{Perez:2015pea}
R.~Navarro~P\'erez, E.~Ruiz~Arriola and J.~Ruiz~de Elvira,
  \emph{{Self-consistent statistical error analysis of $\pi\pi$ scattering}},
  \href{https://doi.org/10.1103/PhysRevD.91.074014}{\emph{Phys. Rev.}
  {\bfseries D91} (2015) 074014},
  [\href{https://arxiv.org/abs/1502.03361}{{\ttfamily 1502.03361}}].

\bibitem{Perez:2013jpa}
R.~Navarro~P\'erez, J.~E. Amaro and E.~Ruiz~Arriola, \emph{{Coarse-grained
  potential analysis of neutron-proton and proton-proton scattering below the
  pion production threshold}},
  \href{https://doi.org/10.1103/PhysRevC.88.064002,
  10.1103/PhysRevC.91.029901}{\emph{Phys. Rev.} {\bfseries C88} (2013) 064002},
  [\href{https://arxiv.org/abs/1310.2536}{{\ttfamily 1310.2536}}].

\bibitem{RuizArriola:2016sbf}
E.~Ruiz~Arriola, J.~E. Amaro and R.~Navarro~P\'erez, \emph{{The falsification
  of Chiral Nuclear Forces}},
  \href{https://doi.org/10.1051/epjconf/201713709006}{\emph{EPJ Web Conf.}
  {\bfseries 137} (2017) 09006},
  [\href{https://arxiv.org/abs/1611.02607}{{\ttfamily 1611.02607}}].

\bibitem{Reinert:2017usi}
P.~Reinert, H.~Krebs and E.~Epelbaum, \emph{{Semilocal momentum-space
  regularized chiral two-nucleon potentials up to fifth order}},
  \href{https://doi.org/10.1140/epja/i2018-12516-4}{\emph{Eur. Phys. J.}
  {\bfseries A54} (2018) 86},
  [\href{https://arxiv.org/abs/1711.08821}{{\ttfamily 1711.08821}}].

\bibitem{Entem:2017gor}
D.~R. Entem, R.~Machleidt and Y.~Nosyk, \emph{{High-quality two-nucleon
  potentials up to fifth order of the chiral expansion}},
  \href{https://doi.org/10.1103/PhysRevC.96.024004}{\emph{Phys. Rev.}
  {\bfseries C96} (2017) 024004},
  [\href{https://arxiv.org/abs/1703.05454}{{\ttfamily 1703.05454}}].

\bibitem{Salpeter:1951sz}
E.~E. Salpeter and H.~A. Bethe, \emph{{A Relativistic equation for bound state
  problems}}, \href{https://doi.org/10.1103/PhysRev.84.1232}{\emph{Phys. Rev.}
  {\bfseries 84} (1951) 1232--1242}.

\bibitem{Carbonell:1998rj}
J.~Carbonell, B.~Desplanques, V.~A. Karmanov and J.~F. Mathiot,
  \emph{{Explicitly covariant light front dynamics and relativistic few body
  systems}}, \href{https://doi.org/10.1016/S0370-1573(97)00090-2}{\emph{Phys.
  Rept.} {\bfseries 300} (1998) 215--347},
  [\href{https://arxiv.org/abs/nucl-th/9804029}{{\ttfamily nucl-th/9804029}}].

\bibitem{Nieves:1999bx}
J.~Nieves and E.~Ruiz~Arriola, \emph{{Bethe-Salpeter approach for unitarized
  chiral perturbation theory}}, {\emph{Nucl. Phys.} {\bfseries A679} }.

\bibitem{Arriola:2016pnw}
E.~Ruiz~Arriola, \emph{{Some Three-body force cancellations in Chiral
  Lagrangians}},  \href{https://arxiv.org/abs/1606.07535}{{\ttfamily
  1606.07535}}.

\bibitem{namyslowski1967relativistic}
J.~Namyslowski, \emph{Relativistic, 3-dimensional, 2-body integral equations.
  on-shell and off-shell formalisms}, {\emph{Physical Review} {\bfseries 160}
  (1967) 1522}.

\bibitem{Lutz:2015lca}
M.~F.~M. Lutz, E.~E. Kolomeitsev and C.~L. Korpa, \emph{{Spectral
  representation for u- and t-channel exchange processes in a partial-wave
  decomposition}},
  \href{https://doi.org/10.1103/PhysRevD.92.016003}{\emph{Phys. Rev.}
  {\bfseries D92} (2015) 016003},
  [\href{https://arxiv.org/abs/1506.02375}{{\ttfamily 1506.02375}}].

\bibitem{Allen:2000xy}
T.~W. Allen, G.~L. Payne and W.~N. Polyzou, \emph{{Comparison of relativistic
  nucleon-nucleon interactions}},
  \href{https://doi.org/10.1103/PhysRevC.62.054002}{\emph{Phys. Rev.}
  {\bfseries C62} (2000) 054002},
  [\href{https://arxiv.org/abs/nucl-th/0005062}{{\ttfamily nucl-th/0005062}}].

\bibitem{RuizSimo:2017anp}
I.~R. Simo, J.~E. Amaro, E.~Ruiz~Arriola and R.~Navarro~Perez, \emph{{Low
  energy peripheral scaling in nucleon-nucleon scattering and uncertainty
  quantification}}, \href{https://doi.org/10.1088/1361-6471/aaabd2}{\emph{J.
  Phys.} {\bfseries G45} (2018) 035107},
  [\href{https://arxiv.org/abs/1705.06522}{{\ttfamily 1705.06522}}].

\bibitem{Fernandez-Soler:2017kfu}
P.~Fernandez-Soler and E.~Ruiz~Arriola, \emph{{Coarse graining of NN inelastic
  interactions up to 3 GeV: Repulsive versus structural core}},
  \href{https://doi.org/10.1103/PhysRevC.96.014004}{\emph{Phys. Rev.}
  {\bfseries C96} (2017) 014004},
  [\href{https://arxiv.org/abs/1705.06093}{{\ttfamily 1705.06093}}].

\bibitem{Perez:2013cza}
R.~Navarro~Perez, J.~E. Amaro and E.~Ruiz~Arriola, \emph{{Partial Wave Analysis
  of Chiral NN Interactions}},
  \href{https://doi.org/10.1007/s00601-014-0817-3}{\emph{Few Body Syst.}
  {\bfseries 55} (2014) 983--987},
  [\href{https://arxiv.org/abs/1310.8167}{{\ttfamily 1310.8167}}].

\bibitem{RuizdeElvira:2018hsv}
J.~Ruiz~de Elvira and E.~Ruiz~Arriola, \emph{{Coarse graining $\pi \pi $
  scattering}},
  \href{https://doi.org/10.1140/epjc/s10052-018-6342-7}{\emph{Eur. Phys. J.}
  {\bfseries C78} (2018) 878},
  [\href{https://arxiv.org/abs/1807.10837}{{\ttfamily 1807.10837}}].

\bibitem{Alarcon2019}
J.~M. Alarc\'on, J.~Ruiz~de Elvira and E.~Ruiz~Arriola, \emph{{Coarse graining
  $\pi N$ scattering}}, {\emph{In preparation} }.

\bibitem{Perez:2013oba}
R.~Navarro~P\'erez, J.~E. Amaro and E.~Ruiz~Arriola, \emph{{Coarse grained NN
  potential with Chiral Two Pion Exchange}},
  \href{https://doi.org/10.1103/PhysRevC.89.024004}{\emph{Phys. Rev.}
  {\bfseries C89} (2014) 024004},
  [\href{https://arxiv.org/abs/1310.6972}{{\ttfamily 1310.6972}}].

\bibitem{Perez:2014bua}
R.~Navarro~P\'erez, J.~E. Amaro and E.~Ruiz~Arriola, \emph{{Low energy chiral
  two pion exchange potential with statistical uncertainties}},
  \href{https://doi.org/10.1103/PhysRevC.91.054002}{\emph{Phys. Rev.}
  {\bfseries C91} (2015) 054002},
  [\href{https://arxiv.org/abs/1411.1212}{{\ttfamily 1411.1212}}].

\end{thebibliography}

\providecommand{\href}[2]{#2}\begingroup\raggedright\endgroup

\end{document}